\documentclass[english]{cccconf}
\usepackage[comma,numbers,square,sort&compress]{natbib}
\usepackage{epstopdf}
\begin{document}

\title{Life detection strategy based on infrared vision and ultra-wideband radar data fusion}

\author{li.yin\aref{siat,ris,hmris}, ym.zhou\aref{siat,ris,hmris}}



%

\affiliation[siat]{ Shenzhen Institutes of Advanced Technology(SIAT), Chinese Academy of Sciences, Shenzhen 518055, P. R. China }
\affiliation[ris]{Guangdong Provincial Key Lab of Robotics and Intelligent System, SIAT, Chinese Academy of Sciences }
\affiliation[hmris]{ Key Laboratory of Human-Machine Intelligence Synergic Systems, SIAT, Chinese Academy of Sciences \email{ li.yin@siat.ac.cn, ym.zhou@siat.ac.cn}}

\maketitle

\begin{abstract}

The life detection method based on a single type of information source cannot meet the requirement of post-earthquake rescue due to its limitations in different scenes and bad robustness in life detection. This paper proposes a method based on deep neural network for multi-sensor decision-level fusion which concludes Convolutional Neural Network and Long Short Term Memory neural network (CNN+LSTM). Firstly, we calculate the value of the life detection probability of each sensor with various methods in the same scene simultaneously, which will be gathered to make samples for inputs of the deep neural network. Then we use Convolutional Neural Network (CNN) to extract the distribution characteristics of the spatial domain from inputs which is the two-channel combination of the probability values and the smoothing probability values of each life detection sensor respectively. Furthermore, the sequence time relationship of the outputs from the last layers will be analyzed with Long Short Term Memory (LSTM) layers, then we concatenate the results from three branches of LSTM layers. Finally, two sets of LSTM neural networks that is different from the previous layers are used to integrate the three branches of the features, and the results of the two classifications are output using the fully connected network with Binary Cross Entropy (BEC) loss function. Therefore, the classification results of the life detection can be concluded accurately with the proposed algorithm.

\end{abstract}

\keywords{Multi-sensor fusion, Life detection, Deep learning, CNN+LSTM}


\section{Introduction}

Studies have shown that China has entered a new period of active earthquakes and has always faced the threat of a strong earthquake\cite{earthquake}. How to improve the search efficiency and reduce the damage of survivors in the earthquake, and how to determine the existence and location of people in the ruins is of major challenging\cite{existence and location}.

Traditional life detection uses a single sensor. Ultra Wide Band (UWB) life signal electromagnetic detection is an advanced life detection technology, which uses the reflection principle of electromagnetic waves to detect the micro-motion caused by human breathing and heartbeat\cite{uwbjiance}. Infrared video life detection is used to detect the living body state at night or under high noise, by collecting temperature distribution of the living body\cite{hongwai}. The weak acoustic wave detection is used to detect and identify weak sound signals of a living body, which always perform auxiliary rescue for life detection\cite{sound}. Each sensor has its own advantages and disadvantages, and plays an important role in the life detection\cite{multi sensor review}.

The current technical disadvantage is that a single sensor cannot meet the life detection requirements of today's complex scenes. A single sensor relies heavily on the environment and is subject to serious environmental interference. It is important to fuse different sensors to improve accuracy of the search, when we choose infrared video or acoustic waves and other methods for life detection. For example, in places where electromagnetic interference is relatively strong, auxiliary detection can be performed by means of audio signals, and UWB radar can be used for detection in the case of complex building coverage or relatively large thermal noise\cite{advantage multi-sensor fusion}.

A life detection method based on multi-sensors data fusion such as infrared and UWB radar is proposed, whose robustness and accuracy are improved. At the same time, the discriminant model proposed in this paper based on the probability of life detection by multi-sensor fusion can be applied to the mobile life detection platform more conveniently. For example, reducing the hardware volume can be installed on the drone, and the drone can give life detection results through the wireless signal based on the life detection probability and the location of the drone, which will give a quick and accurate guide to the first aid staff.

\section{Life detection sensors}

This paper proposes a multi-sensor based life detection method and strategy. The sensor fusion part uses a three-input and single-output neural network to predict the probability of life existence, whose inputs of the neural network is the probability value of life detection for each sensor, including the life detection probability value of UWB radar, infrared video, and weak acoustic wave. Among them, the processing difficulty is UWB radar life detection, and the UWB radar adopts Principal Component Analysis (PCA) dimensionality reduction and CNN+LSTM feature extraction method to obtain the probability of life detection.

\subsection{Ultra wide band radar sensor}

The traditional UWB radar life detection method is to coherently process the original signal and the target echo signal to obtain the life detection result. For the vital signal, the radar echo signal can be regarded as UWB radar original signal modulated by the human micro-motion, the surrounding environment and other clutter signals. It is assumed that the channel impulse response of the UWB radar can be expressed as:

\begin{equation}
  \label{eq1}
  h(\tau ,t) = {a_1}\delta (\tau  - {\tau _1}(t)) + \sum \limits_i {{a_i}} \delta (\tau  - {\tau _i}(t))
\end{equation}

$t$ is the fast time sampling signal of UWB radar, which can be used to represent distance information. $\tau$ is the slow time sampling signal of UWB radar, whose reciprocal is the Pulse Repetition Frequency (PRF). Assuming that the radar transmits a signal $p(\tau)$, the echo signal is expressed as:

\begin{equation}
  \label{eq2}
  R(\tau ,t) = p(\tau )*h(\tau ,t) = {a_1}p(\tau  - {\tau _1}(t)) + \sum\limits_i {{a_i}p(\tau  - {\tau _i})}
\end{equation}

\begin{equation}
  \label{eq3}
  {\rm{R}}(m,n) = r(m{T_s},n{T_f})
\end{equation}

\begin{equation}
  \label{eq4}
  {\rm{R}}(m,n) = {a_1}p(m{T_s} - {\tau _1}(n{T_f})) + \sum\limits_i {{a_i}p(mT - {\tau _i})}
\end{equation}

Where ${T_s}$ and ${T_f}$ are the sampling interval in slow frequency and fast frequency, $m = 0,1,2,...,M - 1; n = 0,1,2,...,N - 1$, and $R(m,n)$ is the radar echo matrix that carries vital signs information.

The radar signal is a non-stationary nonlinear signal, so that the characteristics of its frequency region cannot be effectively separated with wavelet transform. PCA can separate the components of the echo signal, suppress the clutter and improve the signal matrix\cite{uwb pca}. SIGNAL-NOISE RATIO (SNR) can be decomposed into subspaces of vectors with the largest variance by the PCA method, so that the radar signals with high SNR can be reconstructed.

Empirical Mode Decomposition (EMD) is a time-scale signal decomposition method based on the time scale of the data itself which obtain a series of Intrinsic Mode Function (IMF) containing local features of different time scale signals\cite{uwb formula cite}. EMD algorithm can also avoid the instantaneous frequency fluctuation caused by the signal asymmetry while defining the instantaneous frequency. This is because each IMF component requires the following two conditions:

\begin{itemize}
  \item The data is distributed the entire time space, and the number of extreme points (including the maximum value and the minimum value) is equal to or at most one different from the zero crossing point, and the minimum value point and the maximum value point need to be enveloped.
  \item At any moment time, the mean of the upper and lower envelopes consisting of local extreme points is zero.
\end{itemize}

The EMD decomposition on the signal is also called a screening process which has two functions. It can remove the superimposed waveform and make the data more symmetrical. The specific steps are as follows:

\begin{itemize}
  \item To find all the maxima and minima points of s, and use the fitted values to get the upper and lower envelope curves $\max$ and $\min$ respectively.
  \item When calculate the mean value $m_1$, ${m_1} = ({l_{\max }} + {l_{\min }})/2 $, the data $s(t)$ subtract the mean value, ${h_1} = s(t) - {m_1}$, we can obtain ${h_1}$. If the obtained $h_1$ can satisfy the IMF constraint, then the mark $IM{F_1} = {h_1}$.
  \item Continue ${d_1} = s(t) - IM{F_1}$.
  \item Repeat the above steps until $d_n$ can't be decomposed, ${d_n}$ is the trend of $s(t)$, the data reconstructed after decomposition can be expressed as: $x(t) = {d_n} + \sum\limits_i^n {IM{F_i}}$.
\end{itemize}

After the same signal is decomposed frequency vector with EMD, the IMF component is obtained sequentially from high frequency to low frequency. Each IMF component represents different vital signs, and the center hop frequency is higher than the respiratory signal frequency, so it can make predictions about the life detection based on the frequency of the IMF component.

However, the cost of EMD decomposition is computationally intensive. Moreover, it is more complicated to reconstruct the vital sign signal and select the characteristics of the target echo signal such as respiration, heartbeat. This paper proposes a method for extracting features by CNN+LSTM on the radar signal. CNN can easily extract the important features of the weak signals and enhance the lack of feature mode in feature extraction, with the echo signal preprocessed with PCA. LSTM network can enhance the feature relationship between time scales and reduce the instantaneous frequency fluctuations without reconstruction of vital signs. The disadvantage is that the signal can't be symmetrical like the EMD, and also CNN+LSTM method does not highly rely on the symmetry of the signal by using the affine transformation between the neural networks.

\begin{figure*}
  \centering
  \includegraphics[width=\hsize]{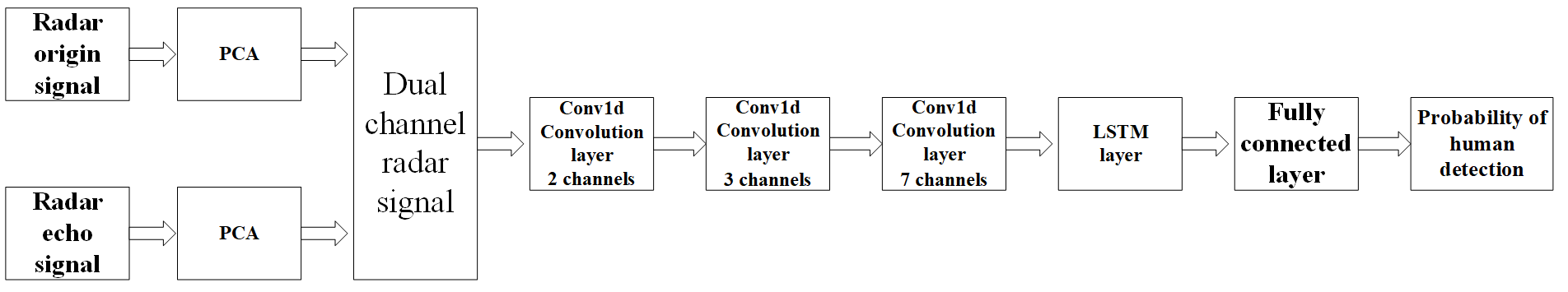}
  \caption{the UWB radar based life detection}
  \label{fig1}
\end{figure*}

We use the PCA to preprocess the data to suppress the clutter component in the echo signal\cite{uwb radar} with the same operation in the original UWB signal. In order to obtain the neural network input samples, we combine the original wave signal and the echo signal into a two-channel radar wave signal. CNN is used to extract two-channel signal feature, and LSTM is used to analyze the correlation between the two-channel signal time scales. Then the neural network can determine whether there is a living body by a supervised classifier with a probability value of the life detection. The design of proposed neural network is shown in Fig~\ref{fig1}.

\subsection{Infrared video sensor}

In this paper we use the infrared background difference method, the target feature extraction method, and Support Vector Machine (SVM) classification method to obtain the probability value of the life detection. Human target detection is mainly divided into Region Of Interest (ROI) segmentation process and classification detection process. The infrared image sequence ROI segmentation methods mainly include optical flow method, frame difference method and background difference method\cite{infrad cut}. The background difference method uses the current frame image to differentiate from the reference background. Common methods for classification detection of human targets include template-based matching methods, recognition methods based on target motion information, and methods based on target feature extraction classification such as SVM.

The adaptive updating GMM algorithm is used to segment the ROI in the infrared image sequence to obtain the human candidate target. Then, we classify the AOE-HOG feature that is extracted from the candidate target by SVM which can improve the real-time and accuracy to certain extent.

\begin{figure}
  \centering
  \includegraphics[width=\hsize]{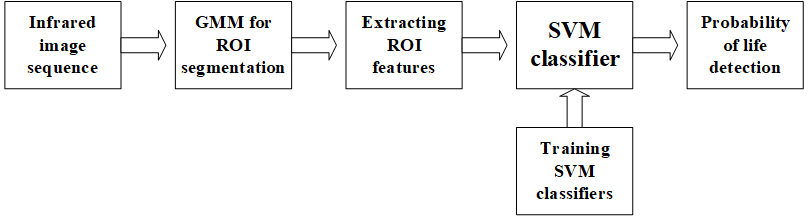}
  \caption{the infrared video based life detection}
  \label{fig2}
\end{figure}

SVM\cite{infrad svm} is a supervised machine learning method. The classification of the target area is realized by constructing a classification interface. That is, a series of positive and negative training samples are used to train and optimise the classifier. The infrared video sensor processing is shown in Fig~\ref{fig2}.

In this paper, when analyzing the infrared video, we can obtain the probability value of the living body from the network, use a classifier at the end of neural network which is good for multi-sensor fusion\cite{infrad classification}.

\subsection{Acoustic wave sensor}

Sound signal can be used for life detection, which can be used to analyze various useful acoustic waves such as low frequency interval tap wave and physiological signals issued of trapped people such as weak respiration and heartbeat. The life detection can be achieved by these two kinds of vital life signal.

We first adopt wavelet transform method to filter the noise of the signals by threshold for the original signal. Then, the wavelet inverse transform is used to obtain the required effective signal with little noise, which is useful for the feature extraction and classification. We use Independent Component Analysis (ICA) to separate the spectrum because the mixed multi-source signals are often difficult to be separated, since ICA is a multi-channel blind source separation method developed by blind source separation technology\cite{acoustic ica}. The original signals must be independent from each other to satisfy the requirement of neural network in signal process\cite{acoustic signal process}.

The acoustic wave life detection is as the same as the UWB radar life detection which also use correlation analysis principle to analyze the original signal. The correlation function is defined as,

\begin{equation}
  \label{eq5}
  {r_{xy}}(m) = \sum\limits_{n =  - \infty }^\infty  {x(n)y(n + m)}
\end{equation}

Where $x(n)$ and $y(n)$ are mutual correlation, ${r_{xy}}(m)$ is the value at time m, which can determine whether there is life by comparing the similarity of each ICA signals. If multiple separated signals are all similar, then there is no life, on the contrary there may be life whose accuracy is not high in life detection.

In this paper, convolutional neural network can replace the multi-channel independent component analysis to calculate the possibility of the existence of the living body which has high accuracy and low complexity. The specific process is shown in Fig~\ref{fig3}:

\begin{figure}
  \centering
  \includegraphics[width=\hsize]{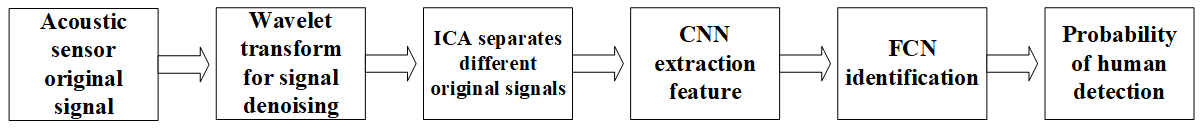}
  \caption{the acoustic wave based life detection}
  \label{fig3}
\end{figure}

\section{Decision-level sensor information fusion}

In this paper, a life detection method and strategy based on multi-sensor fusion is proposed by comparing and analysing the basic theory of Dempster-Shafer (D-S) with neural network method, which uses a three-input, single-output and two-class neural network to predict the probability of a living body. The three sensors will process the signals separately to get the probability value of the life detection of a single sensor before entering the decision level data fusion. This neural network with CNN+LSTM can combine the advantages of multiple sensors to achieve the optimal state of the sensor, and compensate the shortcomings of the D-S evidence theory in the decision-level fusion fixed rule, and can be extended to multiple sensors with high scalability\cite{method multi sensor}.

The traditional decision-level data fusion method of life detection generally uses the D-S evidence theory to conclude, so that the most reliable life detection sensor can be selected. The D-S synthesis rule combines multiple topics, such as predictions of different people or predictions of different sensors. By using D-S evidence theory for decision-level fusion, the priori data is more intuitive and more integrated than that of the probabilistic reasoning to obtain, which can be combined with a variety of data, especially for heterogeneous heterogeneous data fusion\cite{D-S method}. However, the shortcoming is that the evidence synthesis rule does not have strong theoretical support, and its rationality and effectiveness are still controversial. Furthermore there is an exponential explosion problem in calculation. The artificial neural network can learn the distribution of non-standard probability. Even if there is correlation between the original data, the hidden distribution of the original data can be reflected by mapping between the neural network neurons. The neural network has strong nonlinearity\cite{cnn lstm}, and it can adapt to the irregularity and robustness in decision-level reasoning.

In this paper, the proposed sensor fusion based on CNN+LSTM neural network has high performance and robustness\cite{multi sensor cnn lstm}. The decision-level fusion method is used to fuse the life detection result from three sensors to a probability, which gives the most reasonable life detection estimate. In order to enhance the robustness of the decision-level neural network, we select the G time length decision information of the three sensor processing results and combine them into a vector of length 2*G\cite{multi sensor g vector}. Then the problem becomes a long-term sequence classification problem whose time width is G. We can use the three-way parallel network whose first layer is convolutional network to process the input which is the original and smooth signal. Then output of the three-way network is concatenate, and multiple LSTM networks of different sizes are used to perform inter-sequence analysis on the concatenate data. Finally we use a full connection network output to predict probability value with BEC loss function which is suitable for the two-class network. The accuracy will depend on the parameters of network, and the actual tuning of the network is also quite important.

The input signal of each channel is concatenated into a two-dimensional vector using the original signal and the smoothed probability value sequence. Then convolutional neural network is used for feature extraction. If the feature from original signals whose envelope of the signal is sharp, the value of neural network is relatively unstable. Besides, the experiment also proves that if smoothing is not performed, the neural network will not learn anything but average probability value. On the contrary the neural network can not only learn its envelope information, but also preserve the original distribution of the original signal waveform, which can make prediction of probability values for multiple sensors reasonable.

\begin{figure*}
  \centering
  \includegraphics[width=\hsize]{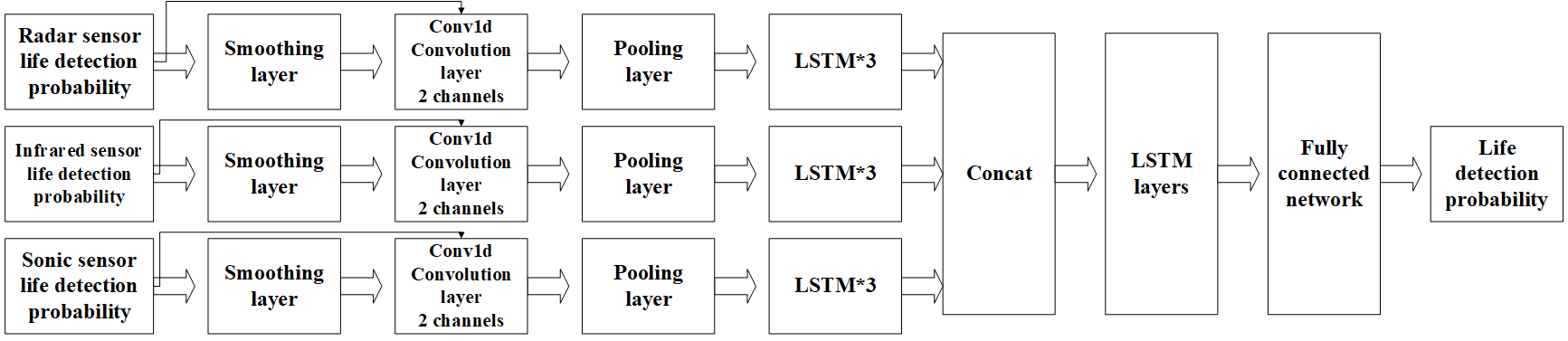}
  \caption{decision-level data fusion neural network}
  \label{fig4}
\end{figure*}

The advantage of the proposed network is that the sensor can be adjusted according to the results of its traditional methods, in the BEC loss function of formula~\ref{loss functioon} below. $y_n$ is the true tag value, which can be 0 or 1, and $x_n$ is the predicted value under the sample. $w_n$ is the weight that can be adjusted. The weight value of the sensor will be lowered when it is classified at the end, and the weight value is not affected by Back Propagation (BP). This function can be realized by configuring a weight list. The convolutional neural networks which is to extract a time series with a certain width is a general method for processing sequence features. The convolutional neural network extracts the distribution characteristics of the sequence values, which can reflect the nature of the data processed by the sensor. The LSTM network can learn the characteristics, and the specific network structure is shown in Fig~\ref{fig4}.

\begin{equation}
  \label{loss functioon}
  {l_n} =  - {w_n}[{y_n} \times \log {x_n} + (1 - {y_n}) \times \log (1 - {x_n})]
\end{equation}

The parameters of the fusion algorithm is particularly important when the classification results of the two methods are combined using the decision-level fusion method. In the estimation of variable length sequences, the choice of time width is more important, which directly determines the validity of the features of the convolutional network. Moreover, the parameters of different sensors are also different. Since it is a decision-level fusion\cite{multi sensor fusion}, the features in the selection of time width may not be obvious. However, the network has a certain time width range which is concluded by personal. The time width is summarized in the experiments based on the characteristics of the network, generally between 64-bit width and 128-bit width.

\section{Experiment}

The experimental UWB radar data in this paper comes from MATLAB generated radar data simulator. We adopt a total 1000 samples of randomly generated waveform data with each 1s length. 1000 infrared images of human bodies with different human postures can be captured by a passive thermal infrared camera. The weak acoustic signal can also use a weak audio collector to collect the weak sound of the tapping of different objects with 1000 samples in 1s length.

Through multiple neural network training, the single variable method is used to set different parameters, including the window width of the smoothing layer, the parameters of the lost layer, the size of the convolutional layer convolution kernel, the number of layers of the LSTM, the LSTM size of the fusion stage, and the whole connection size, etc. The purpose is to observe the best network state under different parameters. Simultaneously, we use BCE loss function and Adam optimizer to train the neural network with momentum parameter to enhance the training speed. The training uses batch technology to process multiple samples, which can speed up the convergence rate of neural network.

\subsection{Sensor data processing}

Data preprocessing is performed using the methods described herein respectively. The transmitted signal of the UWB radar signal and the target echo signal can form a dual channel, which can be classified by the neural network. AOE-HOG features is extracted of the infrared video and they are classified to obtain the probability value of the infrared video with SVM. The acoustic waves uses are amplified to separate the feature by wavelet transform and ICA method. Then the features are extracted by a convolutional neural network to obtain the probability value of the final life detection. The obtained sequence of the probability values are smoothed by the window with width H, so as to concatenate the smoothed probability value sequence and the sequence of probability values into a two-dimensional vector.

\subsection{Result $\&$ Analysis}

\begin{figure}
  \centering
  \includegraphics[width=\hsize]{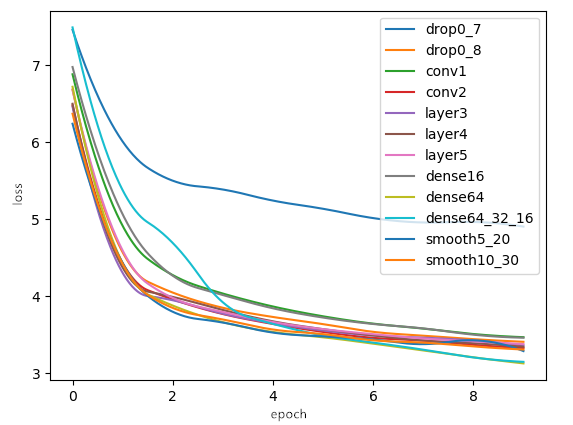}
  \caption{Train loss of different neural network parameters}
  \label{train loss}
\end{figure}

\begin{figure}
  \centering
  \includegraphics[width=\hsize]{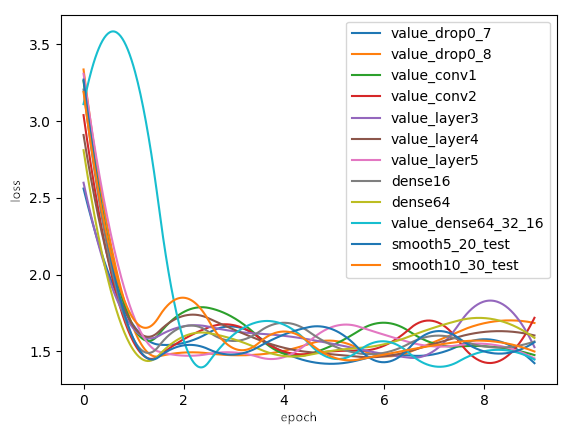}
  \caption{Test loss of different neural network parameters}
  \label{test loss}
\end{figure}

Fig~\ref{train loss} and Fig~\ref{test loss} shows the results of the network components with different network parameters. We set the network structure with the smooth layer width of 5, Conv1d kernel size of 3, Drop out of 0.8, and LSTM layer of 3. After concatenating the result of the sensors, two LSTMs are used to output signals to a fully connected network, which is structure of dense$64\_32\_16$. In order to compare the impact of different parameters on our proposed network, we set the number of layer3, layer4, layer5 as 3, 4 and 5 which means LSTM layers before concat the data of the sensors, the convolution kernel size conv1, conv2 as 1, 3, the dropout rate $drop0\_7, drop0\_8 as 0.7\&0.8$, the smooth window size smooth5, smooth10 as $5\&10$, the loss function as MSE, the optimizer as Adam. The experiment shows that the network structure is more reasonable according to the results of different parameter shown in Fig~\ref{train loss} and Fig~\ref{test loss}.

{\bfseries Data preprocessing:} The probability value of each sensor output is taken as the sample sequence by taking the first G values of the data with G width, and the samples are sequentially backward. For the sequence of length 1000, (1000-G) 2*G vector sizes can be taken. The sample sequence can be used for subsequent convolution calculation. By taking the sequence value of G width as a sample, the relationship between sequences near the predicted sample value can be well analyzed, which is consistent with the characteristics of convolutional neural network and LSTM neural network.

{\bfseries Training:} This experiment uses dual Tesla K80 GPU as the experimental platform, and it takes one hour to complete the experiment on the Pytorch environment which contains epoch 20, batch 16. The probability value vector G*2 of the three sensors is used, where G is the step size in the sequence of probability values. After 10 epoches of training, the neural network will slowly reach a convergence state, and also it is basically the fine-tuning phase of the network after 10 epoches. After each epoch of training is completed, the entire data sequence is re-scrambled, and the gradient is reduced by batch technology, so that the characteristics of the data can be fully learned.

{\bfseries Feature extraction:} The neural network can extract the distribution characteristics of the two-channel data by a convolutional network. Adopting the form of dual channel, it can not only learn the envelope distribution of the smoothed data, but also learn the detailed distribution of the data before smoothing, which can generate a feature map with distributed features by integrating the features map of the two aspects. The obtained feature map is input into the LSTM*3 neural network, whose hidden state space size is G, to learn the characteristics between the time domains of the data and output the time domain feature map.

{\bfseries Feature synthesis and probability output:} The three time domain feature maps output above are concatenated together and passed through the two LSTM networks. The LSTM has a hidden state space of 3*G. The first network can integrate the time domain characteristics of the three feature maps, and the second LSTM. The network only outputs the last state vector (2*G). The state output of the last output is output through the fully connected network. The fully connected network adopts a three-layer architecture with widths of 2*G, G, and 0.5*G. We use the BEC loss function which is more suitable for the two classification operation. The probability value of the fully connected network output can be used for the two classification operation. At the same time, the advantage of the neural network is that the back propagation algorithm can be used to adjust the weight of the whole network. The weighting of BP makes fusion more intelligent.

\begin{figure}
  \centering
  \includegraphics[width=\hsize]{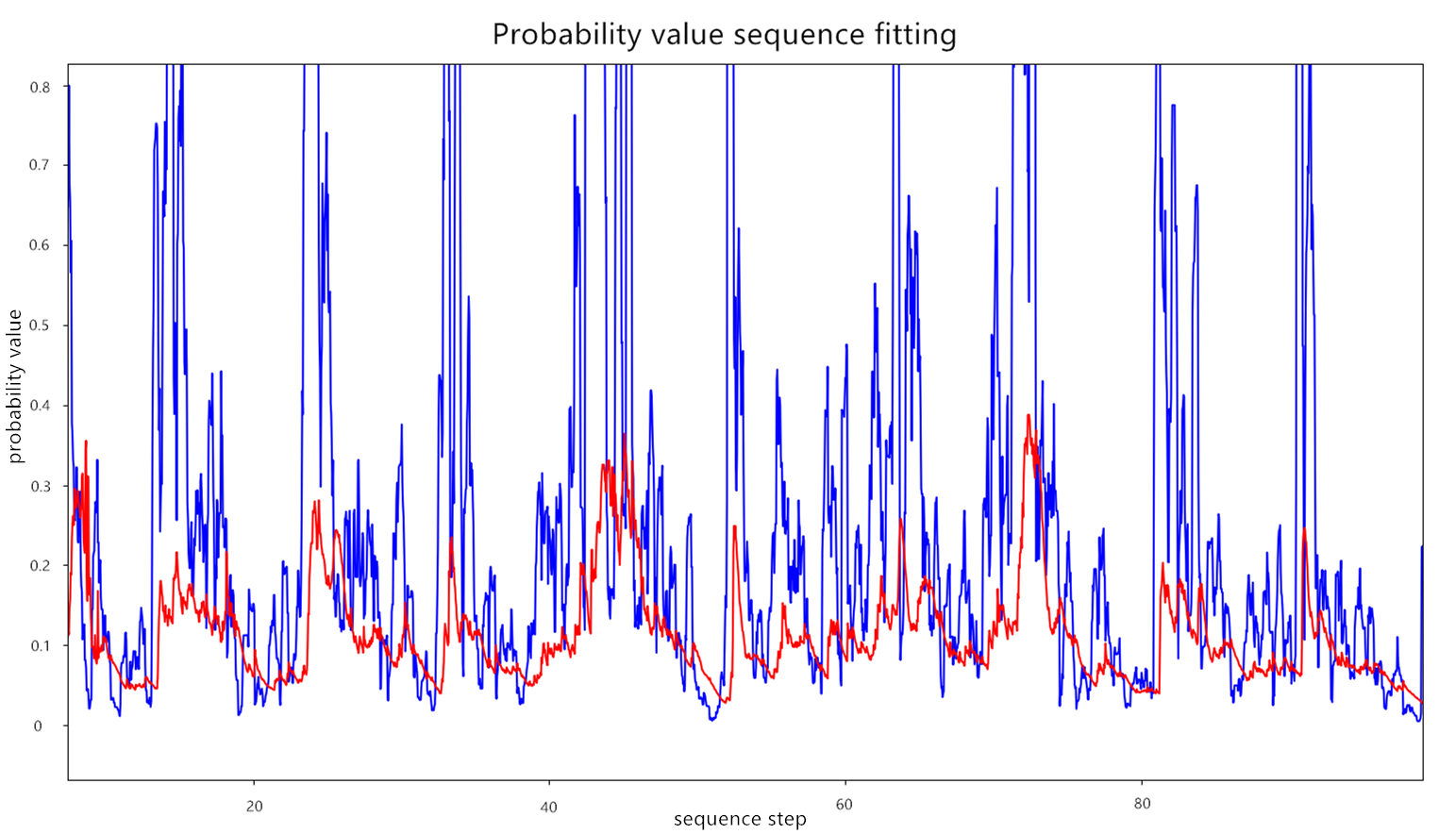}
  \caption{probability value sequence fitting}
  \label{sequence gragh}
\end{figure}

{\bfseries Evaluation metrics:} A comparison chart is drawn between the predicted value and the real value shown in Fig~\ref{sequence gragh}. The red line is the continuous waveform of the predicted value, and the green line is the waveform of the ground truth value. It can be seen from the waveform comparison that the neural network basically fits the waveform of the sample data. The characteristics of the envelope, as well as the detailed trend of the data, basically meet the requirements of multi-sensor data fitting.

\section{Conclusion}

In this paper, the CNN+LSTM neural network model is proposed for decision-level sensor data fusion. The sensor credibility concept is introduced into the sensor, applied to the sensor fusion, and the G length sequence vector input neural network is used for processing. The sensor fusion performance and scalability can enhance the robustness of the system. Firstly, CNN+LSTM is applied to UWB radar sensor feature extraction and probability calculation. The reliability of UWB radar life detection is directly calculated by neural network. Then the credibility value is used when entering decision-level neural network input. The smooth and non-smoothed signal input into the dual-channel CNN network can make full use of the temporal and spatial characteristics of the data to regress the input data and to obtain the desired result. It turns out that the result is in line with expectations. Finally, the three sensor confidence values are processed by the CNN+LSTM network to obtain the final fusion life detection probability value, and the value can be used to determine whether there is vital sign information. It can be applied to mobile environments such as search robots and search and rescue unmanned sets.

\balance

\end{document}